# A Simplified Method of the Assessment of Magnetic Anisotropy of Commonly Used Sapphire Substrates in SQUID Magnetometers

**Katarzyna Gas and Maciej Sawicki \***

Institute of Physics, Polish Academy of Sciences, Aleja Lotnikow 32/46, PL-02668 Warsaw, Poland; kgas@ifpan.edu.pl
\* Correspondence: mikes@ifpan.edu.pl

**Abstract:** Solid state wafers are indispensable components in material science as substrates for epitaxial homo- or hetero-structures or carriers for two-dimensional materials. However, a reliable determination of magnetic properties of nanomaterials in volume magnetometry is frequently affected by unexpectedly rich magnetism of these substrates, including significant magnetic anisotropy. Here, we describe a simplified experimental routine of magnetic anisotropy assessment, which we exemplify and validate for epi-ready sapphire wafers from various sources. Both the strength and the sign of magnetic anisotropy is obtained from carefully designed temperature dependent measurements, which mitigate all known pitfalls of volume SQUID magnetometry and are substantially faster than traditional approaches. Our measurements indicate that in all the samples two types of net paramagnetic contributions coexists with diamagnetism. The first one can be as strong as 10% of the base diamagnetism of sapphire [-3.7(1) × $10^{-7}$ emu/gOe], and, when exceeds 2% mark, it exhibits pronounced magnetic anisotropy with the easy axis oriented perpendicularly to the face of c-plane wafers. The other is much weaker but exhibit ferromagnetic-like appearance. These findings form an important message that non-standard magnetism of common substrates can significantly influence the results of precise magnetometry of nanoscale materials and its existence must be taken for granted by both industry and academia.

## 1. Introduction

A high crystalline quality is one of the key properties of the bulk crystals which are to be used as solid state substrates for the growth of (thin) homo- or hetero- epitaxial structures **[1]**, as well as for two dimensional (2D) materials **[2]**. Some of these structures are expected to become game-changers in spintronics or may exhibit interesting magnetic properties from the fundamental point of view **[3]**. Among a few other characterization methods, the integral magnetometry based on commercial superconducting quantum interferometer device (SQUID) magnetometers is employed to establish the magnetic constitution of these structures. This is where the real challenge begins since the magnetic response of these nanostructures in question is frequently dominated by the signal of its bulky companion, particularly at strong magnetic fields.

An ideal substrate or a sample carrier should not introduce any signal during the measurements, but it always does during the integral magnetometry, so it should be of a simple form and easily accountable. This calls for materials of extreme purity, which (preferably) diamagnetic response should be perfectly linear in magnetic field $H$ and independent of the temperature $T$. This is however a great technological challenge and such materials either cannot be synthesized or their price exceeds budgets of typical research projects. This is probably why, ubiquitously, idealized magnetic properties of these substrates are assumed. Namely, their isotropic and linear in $H$ and $T$-independent diamagnetic susceptibility is assumed and subtracted to obtain the searched response from the nanomaterial. Such an approach may lead even to qualitatively false conclusions. Even commercially available Si or GaAs wafers do show disturbing nonlinear temperature and field dependent responses, particularly at cryogenic temperatures [4], yet systematic data are scarce, since it is hard to account for and report effects which forms and magnitudes vary from sample to sample.

Sapphire ($Al_2O_3$) belongs to these crystalline compounds which chemical purity is still under development and is rather inferior to that of Si or GaAs, yet sapphire wafers are frequently chosen as the base for the deposition of nitride family structures and 2D materials for spintronics applications [5–10]. In the context of magnetism-oriented studies sapphire substrates have been widely used to deposit dilute magnetic nitrides in an epitaxial manner, mainly GaN:Fe [11–13], GaN:Gd [14,15] and (Ga,Mn)N [16–20], to name the most prominent ones. The assumption of the ideal magnetic properties of sapphire is on one hand side supported by results of theoretical calculations [21], and on the other by some experimental communication, e.g. [4,22]. However, there exists another body of experimental evidences pointing out that sapphire does exhibit pronounced $H$- and $T$-dependent contributions, e.g. [23,24,16,25]. Unattended, these strongly nonlinear in $T$ and $H$ components will hinder accurate evaluations of the magnetic response of the structures under the study, and may lead even to qualitatively false conclusions. Therefore, either a great experimental care, backed by an adequate experimental routine [26,27], or a dedicated *in situ* compensation approach [25] are needed to properly assess and mitigate this detrimental contributions. It was found that this surplus magnetic response was predominantly caused by Cr present in the bulk of the material [23] or by a surface contamination by Fe [22]. The more recent studies indicate that Er dopant has assumed the dominant role.

It has to be mentioned that the effects caused by contaminations due to handling, deposition processes and further processing can be even stronger than those pointed above. In the first context it has to be noted that sapphire is much harder than most of the tools used around the laboratory, so it acts as an abrasive material eagerly accumulating small contaminations on the tool-crystal contact surfaces [22,28]. In the second context the possibility of magnetic contaminations by residues of metallic glues or substrate backside metallization used to attach and/or thermalize the substrates in the growth chambers has to be brought into attention. Obviously, the magnitude of this spurious magnetic signal varies a great deal, but can easily exceed $10^{-5}$ emu at room temperature. These contaminations exhibit an overall sigmoidal or Langevin-like shape of their magnetization curves, which saturate swiftly at about 5 kOe [29–31]. The size of this effect makes it comparable to the magnitude of magnetic responses expected in magnetometry of nanomaterials, e.g. [3,32–37], and its ferromagnetic-like appearance could falsely constitute a basis to invoke the existence of ferromagnetism (FM) in the investigated nanomaterial.

Yet another important aspect of the nontrivial magnetic properties of the commonly met solid state substrates comes into light when the magnetic anisotropy (MA) of the nanomaterials has to be investigated. Frequently the form and the magnitude of MA provides an indispensable information about the underlying processes in the investigated material. Equally importantly, a detailed knowledge of MA is required to validate some theoretical considerations [38,39]. This necessitates measurements in both in plane orientation of $H$ (the standard one), and in a more experimentally cumbersome, the perpendicular one. This is in particular "a must do" experiment in the case of aniferromagnetic materials, which, particularly in the form of very thin layers are challenging to measure due to their rather weak magnetic response [40,41]. Finally, the evaluation of the magnetic properties has to be performed in the same experimental configuration in which the typical magneto-transport measurements are performed, that is having $H$ oriented perpendicularly to the surface of the flat sample. So, a presence of a significant MA in the substrate would considerably mar the outcome of the investigations if the relevant piece of the substrate had not been properly evaluated in the same experimental conditions.

Therefore, when the orientation-dependent magnetic studies are at stake it is profitable to know whether the substrate material brings its own, intrinsic, anisotropic contribution to the measurements. Having established early enough that the current substrate material exerts too strong magnetic anisotropy, one can either search for more appropriate substrates from other sources or modify the whole experimental approach to give a proper account for the contributions from the substrate. In this report we put forward a simplified method of assessment of the magnitude of magnetic anisotropy of common substrates, considering sapphire as the sole example. It is argued that the method eliminates a large uncertainty brought about by most common artifacts related to the SQUID magnetometric systems and to the arrangement of the measurements. Importantly, thei approach can be

This paper is divided in the following sections. We firstly introduce the material of our study and substantiate the needs for an alternative approach to precise volume magnetometry of nanostructures on bulky substrates. In Section 3 the results of our magnetic characterization of the sapphire specimens are given. They provide a solid justification to the method put forward by us. In Section 4 we detail the method and provide supporting results obtained from the modeling of a system of noninteracting ions exhibiting a sizable single ion magnetic anisotropy. In section 5 we enumerate a possible range of concentrations of magnetic ions which could be responsible for some of the leading paramagnetic contribution to the ideal diamagnetism of sapphire detected in our samples. The conclusions are given in Section 6.

## 2. Materials and Methods

The sapphire specimens investigated here originate from colorless single-side polished α-$Al_2O_3$ (hexagonal) epi-ready 2 inch wafers. They are all *c*-plane, i.e. the surface has the (0001) orientation, as required for epitaxial growth of various nitride structures that we have been focused on in our research **[11,13,18,20,42]** for nearly two decades. The wafers have been acquired from four different vendors. Only for the record we label them as "A" (6 samples), "B", "C", and "D", since in this report we do not aim at any classification of the available material on the market. We concentrate only on the experimental validation of the method of a fast and reliable assessment of MA in *nominally* diamagnetic substrate crystals. The relatively broad origin of the investigated material helps us to generalize our conclusions.

The specimens are either cleft or professionally cut using a wheel saw from 2 inch wafers into approximately 5 × 5 $mm^2$ pieces or into 1.3 × 5 $mm^2$ strips to facilitate orientation dependent studies, as detailed in Appendix A. All the pieces are etched in an ultrasonic bath of HCl for about 15 minutes to remove surface and post-processing contamination. All magnetic measurements are performed between 2 and 350 K and up to 50 kOe in a Quantum Design (QD) MPMS SQUID magnetometer equipped with a low field option (allowing the magnet reset feature). Sufficiently high signal to noise ratio of the results is facilitated by the use of the reciprocating sample option (RSO). For the SQUID measurements the specimens are glued using strongly diluted GE varnish **[43]** to sample holders made of 2 × 0.7 $mm^2$ across and about 19 cm long Si sticks professionally cut from industrial 8 inches wafers **[26]**. The use of Si sticks, employed routinely by the authors for their most sensitive magnetometry studies **[25,33,44]** completely eliminates position-dependent magnetic signals commonly observed when plastic straws are used to fix the sample in SQUIDs **[45,46]**. In the authors view the use of plastic straws, in particular "straight from the box", i.e. without any selection and testing prior to the measurements, is highly inadvisable and should be avoided in precise, high sensitivity magnetometry. In the same context, the usage of Si sticks, as well as quartz paddle sample holders in the QD VSM-SQUID, facilitates also a perfectly reproducible sample positioning, what is a prerequisite condition for the conventional approach to studies of very weak magnetic anisotropies.

Since the purpose of this report is rather practical, we present our results normalized to a standard, approx. 5 × 5 × 0.3 $mm^3$, sapphire piece weighting 30 mg and express them in the experimental units of the magnetic moment $m$ (emu). This will allow an easy and direct comparison of the results presented here with unprocessed results in other studies. Such a 30 mg sapphire specimen exerts about $-10^{-4}$ emu at 10 kOe.

Undoubtedly, the main experimental challenge in volume magnetometry of nanomaterials is caused by the fact that the signal of the bulky substrates is rather strong and increases linearly with $H$ towards $|m| = 10^{-3}$ emu at 70 kOe, whereas that of the nanostructures rarely exceeds $10^{-5}$ emu, and frequently is much weaker. Therefore all magnetic measurements have to be carried out by strictly observing the experimental code **[26]** elaborated to eliminate artifacts and to evade limitations associated with integral SQUID magnetometry **[27]**. Nevertheless, a simple subtraction of the diamagnetic component originating from the sapphire substrate under an assumption that it is linear in $H$ only exposes the resulting data to various artifacts related to the SQUID system and to arrangements of the measurements **[24,47]**. Other approaches are needed. One of the best methods to mitigate these artifacts is a direct *in situ* compensation of the signal of the substrate **[25,48]** or of the carrier **[49]**, another could be the physical separation of the material in study from the troublesome substrates **[31,35,50–52]**. However, the latter approach necessitates to affix the separated material onto another, substantially cleaner material. Anyway, still a sizable disparity between the signal of interest and that of the new carrier may occur. Therefore, one has to resort to an independent assessment of the magnetic anisotropy of this new supporting material. In this paper we put forward a method which simplifies this task. It allows a much faster and more accurate, though relative, assessment and provides an economical way of ranking of the available material(s) intended to be used as the substrates, supports or carriers for studying of the magnetic anisotropy in nanostructures.

## 3. Results and discussion

Figure 1 shows the range of variations $\Delta m(T) = m(T) - m(300 \text{ K})$ from the expected, temperature independent behavior, established for the range of sapphire samples considered here. All these measurements are performed at $H$ = 20 kOe applied in the plane of the samples (that is for $H$ being perpendicular to the wurtzite $c$ axis of $c$-plane sapphire). The reference value, $m$(20 kOe, 300 K), of a $5 \times 5 \times 0.3$ mm$^3$ (~30 mg) piece of sapphire is about $-2.2 \times 10^{-4}$ emu. More exactly, the mass susceptibility $\chi$ of the samples studied here is found to be $\chi_{\text{Sapphire}}$(300 K) = $-3.7(1) \times 10^{-7}$ emu/gOe. This value has been obtained taking the sample-to-SQUID coupling factor $\gamma$ = 0.983 appropriate for $5 \times 5$ mm$^2$ sample investigated in the in-plane configuration **[26]**. We need to modify the results reported by magnetometers by a size and orientation dependent coefficient $\gamma$, which rescales the response of the SQUID software, calculated in the point object approximation, to the real response of a physical objects of given dimensions **[26,53]**. The logarithmic scale of temperature is used only for convenience. We want to expose the low $T$ region, where most of the changes takes place. Indeed, down to 100 K hardly any changes are seen on all $\Delta m(T)$ curves. Curie-type paramagnetic-like deviations develop around 100 K and at 2 K this contribution can be as strong as $2 \times 10^{-5}$ emu. This is actually as much as about 2% of the diamagnetic response of sapphire.

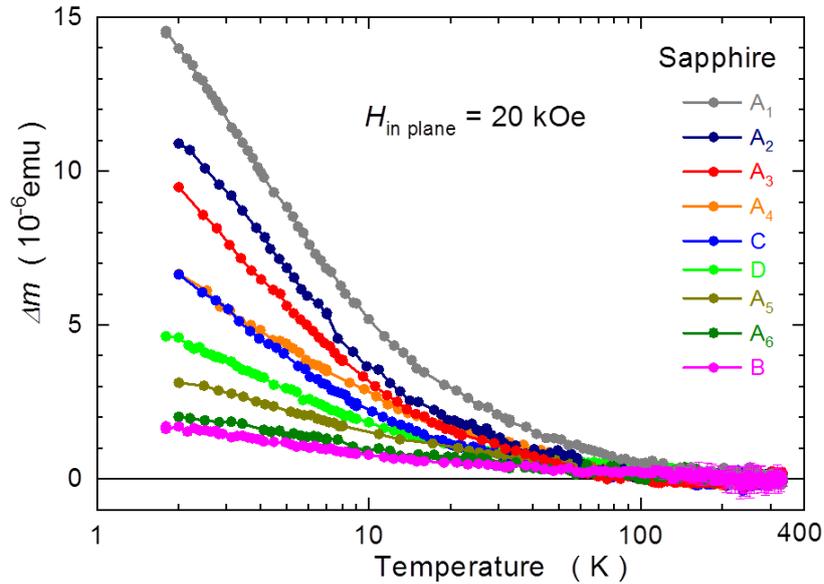

**Figure 1.** Temperature variation of the magnetic moment *m* in various sapphire samples originating from four different vendors, A to D, as indicated in the legend. The results for all the samples are rescaled to correspond to a 30 mg sample and the temperature induced changes *Δm* are calculated with respect to *m*(300 K). Magnetic field *H* of 20 kOe is applied in the plane of the samples, i.e. *H* is perpendicular to the *c* axis of these *c*-plane sapphire specimens. For clarity the error bars are indicated only for sample B.

It has to be pointed out that on the grounds of theoretical considerations no temperature dependence of $\chi$ in sapphire has been expected [21,54]. This stems from the huge energy gap of sapphire, $E_g \cong 9.9$ eV, for which the only *T*-variable contribution to $\chi$, the van Vleck-type paramagnetism, $\chi_{vV}$, determined by the *T*-dependence of the band gap, is practically negligible, since $\chi_{vV} \propto 1/E_g$. Historically, both the *T*-independent properties [4,46] and *T*- and *H*-depended *intrinsic* magnetism in similar sapphire samples [23,24,16,25] were noted. The results collected in Figure 1 clearly contradict the former claims. In neither case *Δm(T)* is *T*-independent. In fact all *Δm(T)* measured by us indicate the existence of a variable in size Curie-like paramagnetic contribution. The overall picture emerges that there is not any "universal sapphire" material. The magnetic response of sapphire substrates available on the market does change from sample to sample and that the differences among them are substantial. This is particularly clearly seen within the range of "A" samples, by far the most numerous species in our study. These samples cover (nearly) the full range of *Δm(T)* observed here. The material from other sources show rather smaller magnitudes of *Δm(T)*, yet still with a sizable distribution. This important finding indicates that commercially available sapphire is not magnetically clean and that researches must take for granted an existence of a certain level of paramagnetic-like contaminants. Lastly, we underline here the most worrying fact of a rather large magnitude of these paramagnetic deviations. If left unaccounted such a behavior of a substrate is sure to mislead any low temperature nano-magnetic studies.

Actually, the magnitudes of the changes of *Δm(T)* reported in Figure 1 tie to a large degree with anisotropic properties of these samples. The temperature dependent data for both the in plane orientation of *H*, $H_{in\ plane}$, and for *H* applied perpendicularly to plane, $H_{perp}$ are exemplified in left panels of Figure 2. Here *Δm(H)* is obtained similarly to the results presented in Figure 1, i.e. the magnetometry data are normalized to 30 mg and additionally rescaled to yield *Δm(T)* = 0 at room temperature both orientations of *H*. A following general pattern emerges. A clear magnetic anisotropy is seen in these samples which exhibit the largest values of *Δm(T)*, say above about $5 \times 10^{-6}$ emu at the lowest temperatures. This is above about 2% of their *m*(20 kOe, 300 K) in relative units. When magnitudes of *Δm(T)* are below this limit no dependency on the orientation of *H* is registered within experi-mental accuracy, as exemplified in panels (e) and (g) of Figure 2. It remains an interesting question if this general picture holds also for another sapphire substrates and whether, or not, a similar threshold value could be identified. An another general result of our scrutiny is that in all cases when we see MA, the positive correction to the *c*-plane sapphire, $\Delta m(T, H)$, is stronger for $H_{perp}$ than for $H_{in\ plane}$. Or $\Delta m(T, H \parallel c) > \Delta m(T, H \perp c)$, when one refers to the crystallographic orientations. We dub this case as the negative MA, since the net paramagnetic contribution to *m* in this by far more frequently exercised orientation in magnetometry, that is with *H* applied in the plane of the substrate is smaller than that when *H* is applied perpendicularly to the face of substrate.

Magnetic measurements performed in the magnetic field domain do confirm the conclusions of the *T*-dependent results, as shown in the right panels of Figure 2. Here, we consider only the nonlinear parts of $m(H)$, $\Delta m(H) = m(H) - \alpha \cdot H$, where $\alpha$ is derived from the slope of $m(H, 300\ K)$. Magnitudes of $\alpha$ are established separately for each sample and each orientation of *H*. Indeed, each measurement run yields its own value of $\alpha$. In practice, $\alpha$ is not directly proportional to $\chi_{sapphire}(300\ K)$. This is because the results of the measurements (the bare numbers reported by the magnetometer) depend also on the magnitude of the orientation-dependent coupling factor $\gamma$. Equally importantly $\alpha$ also contains all other much harder accountable factors which additionally influence the absolute values of *m* reported by the magnetometer. First ones are related to an imperfect sample positioning. These include both the rotational and the radial misalignments of the specimen in SQUID sample chamber with respect to the centerline of the magnetometer. These sources of errors are poorly described in QD technical notes for MPMS systems. One should refer to SQUID VSM technical notes for more details **[55,56]**. The other frequently occurring errors are brought about by the deviations from the ideal performance of the RSO sample transport mechanism. Any of these factors alone can influence the magnitude of the response of the magnetometer by up to 2%. In this very sense $\alpha$ parameters are not sample specific quantities, they are specific to each experimental run as they reflect the current experimental geometry, specimen's (mis-)alignments, and the performance of the mechanical parts of the magnetometer. Therefore, a different value of $\alpha$ is frequently obtained when *the same* sample is re-measured in *nominally* the same arrangements.

The results collected in panels (b), (d), (f), and (h) confirm that the magnetic response of all the sapphire samples studied by us is a paramagnetic-like. However, we can identify two leading patterns. Samples with the large values of $\Delta m(H)$ exhibit a simple Brillouin-like response, yet with a pronounced MA. Similarly to the corresponding *T*-dependent results, the $H_{perp}$ orientation is the easy one. The results shown in the lowest panel (h) exhibit a strong upturn of *m* around $H = 0$, which is followed by a pronounced kink. They do not show any tendency to saturation up to 50 kOe, and MA, if any, is very weak. The different origin of this behavior is also indicated by the fact that the magnitudes of $\Delta m(H)$ in Figure 2 (h) are the smallest. It can be noted that both $\Delta m(H)$ in panel (f) constitute a kind of a border case as they contain both the Brillouin-like and the non-Brillouin-like types of $\Delta m(H)$, with the clear dominance of the former. Actually, the $\Delta m(H)$ reported in panels (f) and (h) are very much alike to those reported in Figure 1 of ref. **[16]**, although the source of these wafers is different.

The case presented in panel (h) illustrates the fact that $\Delta m(H)$ of the epi-ready sapphire substrates may exhibit a deceivingly FM-like character. However, actually no other typical FM features like magnetic hystereses or nonzero remnant moments have been observed in this and all other investigated specimens. This is a substantially different outcome than that of ref. **[22,29]**, where a sizable concentration of Fe or Ti (mostly on the surfaces) was identified to be responsible for strong hystereses of $\Delta m(H)$. This fact underpins the importance of the proper sample cleaning prior to the measurements and highlight the use of non-ferrous tools to handle the wafers and the samples.

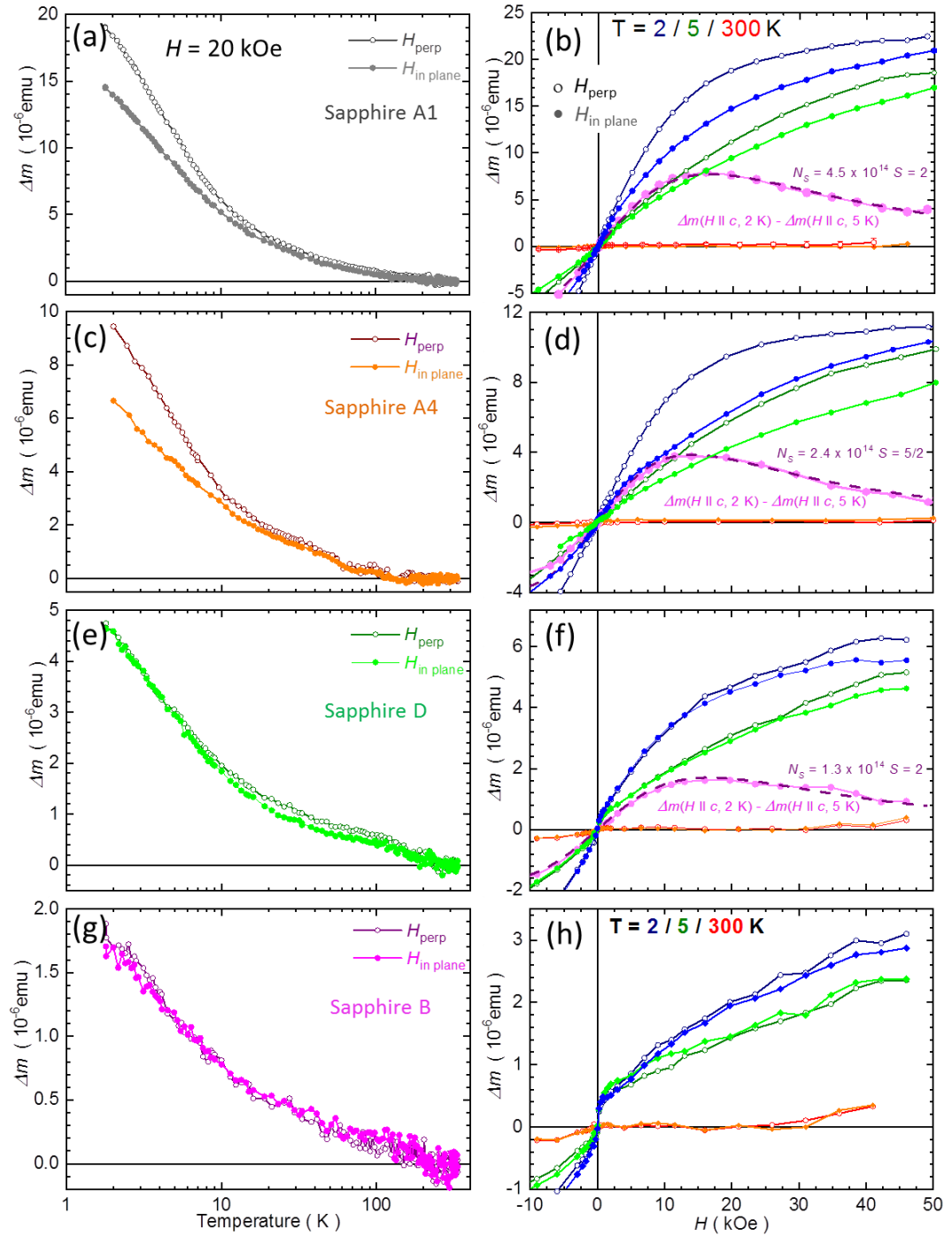

**Figure 2.** Results of the magnetic field $H$ orientation-dependent studies of selected four sapphire specimens in temperature $T$ domain (left panels) and in $H$ domain (right panels). Each pair (**a**)-(**b**), (**c**)-(**d**), (**e**)-(**f**), and (**g**)-(**h**) presents results for one of these samples. Results in $T$ domain are obtained at $H$ = 20 kOe. Here the results are normalized to correspond to a $5 \times 5 \times 0.3$ mm³ (~30 mg) sapphire sample, and $\Delta m(T) = m(T) - m(300\,\text{K})$. The results in $H$ domain are plotted after the application of a liner correction $\alpha \cdot H$, where $\alpha$ is the slope, of $m(H, 300\,\text{K})$; $\Delta m(H) = m(H) - \alpha \cdot H$. Open symbols stand for perpendicular orientation $H$, the closed ones represent $H$ applied in the plane of these $c$-plane sapphire samples. The magenta bullets in $H$-panels (b), (d), and (f) show the difference between $\Delta m(H_{\text{perp}})$ measured at $T$ = 2 and 5 K to evaluate an approximate number $N_S$ of spins $S$ in the sample. The dashed purple lines represent the fit of difference between Brillouin functions for given $S$ at 2 and 5 K. The possible magnitudes of $N_S$ and $S$ yielding the best fit are given in the panels.

Similarly to the results of the *T*-dependent studies, MA is clearly seen in the top two right panels of Figure 2, practically, no anisotropy is seen in the two lower ones. This obvious correspondence between $\Delta m(T)$ and $\Delta m(H)$ forms the basis of our simplified method of the assessment of the magnetic anisotropy in common semiconductor substrates. Instead of typical measurements in the magnetic field domain, which are much longer and more demanding in terms of required precision, reproducibility and signal to noise ratio, we suggest to perform only measurements in the temperature domain, which are generally simpler, less noisy (magnetic field is stable within hours-long *T*-sweeps), and much faster to acquire. The latter means they are also more economical in terms of a higher throughput and lower helium consumption.

Therefore, we postulate that a sufficiently accurate magnetic evaluation of a family of similar semiconductor substrates, including the assessment of the (relative) strength of their magnetic anisotropy can be done only on the account of the temperature dependent measurements. As modelled below, just two temperature sweeps in a relevant temperature range performed in the two required orientations of *H* are sufficient to perform this task. The existence of MA, or its absence, will be clearly seen in the combined plot of $\Delta m(T, H_\perp)$ and $\Delta m(T, H_\parallel)$. If MA is present, the enumeration of the area between these two $\Delta m(T)$ curves will quantify its strength and so will allow the classification or ranking among the specimens (substrates) in the study.

## 4. Description of the method

The key factor underlying the method proposed here is that MA of weakly contaminated solid-state substrates is negligible small at elevated temperatures and develops only at cryogenic ones, that is discarding cases of extrinsic contaminations, e.g. **[29–31].** Under our assumption the high-*T* magnitudes of *m*(*T*) should be the same for both $H_\perp$ and $H_\parallel$, providing us with a very useful normalization feature. This normalization is the key point here, since it allows to mitigate the substantial inaccuracies resulting from different experimental arrangements specific to separate measurements in both orientations of *H* **[26,53,55,56]**. In particular, after the normalization each pair of the *T*-dependent measurements will not be affected by the recent experimental history of the magnetometer. As the result, these two measurements in question do not have to be performed one after another, though this is an advisable option. The normalization also makes the analysis a way easier and increases the confidence level of the results.

In the classical approach the magnitude of MA can be obtained from the integration of the difference between the easy and the hard axis magnetization curves. Here, we suggest to enumerate the area $A_T$ between $\Delta m(T)$ established for two relevant orientations of *H*. $A_T$ does not carry its usual meaning of the magnetic anisotropy energy density, but its magnitude can be used to a very good result as a numerical quantifier to rank a range of specimens among themselves, as it is demonstrated further below.

To substantiate our method we model theoretically an equivalent material system: wurtzite GaN:Mn. Importantly here, Mn substituting Ga in GaN assumes 3+ oxidation state, which due to $L = 2$ and $S = 2$ of its 4 electrons on the *d*-shell exhibits a strong single ion magnetic anisotropy with respect to its wurtzite *c* axis **[16,57]**. GaN:Mn, as sapphire, is also highly insulating. In fact it has been strongly researched recently as an ideal insulating substrate for homoepitaxy of planar high power nitride devices **[58]**. In the paramagnetic concentration range, say below $10^{19}$ cm$^{-3}$, *T*-, *H*-, and orientation-dependent magnetic properties of GaN:Mn are adequately described within the crystal field model of $d^4$ ion in the wurtzite host **[16,39,59,60]**. We follow this approach to compute exemplary *m*(*H*) and *m*(*T*) curves for various strengths of the single ion MA. The latter depends predominantly on the magnitude of the trigonal distortion (along the *c*-axis) and in fact the strength of MA in GaN:Mn has been documented to change by direct magnetic measurements of piezoelectrically modulated structures **[39]**. The strength of MA and the shape of the relevant *m*(*H*) curves also depends strongly on the oc-cupancy of the $d$ level **[38],** what additionally makes the results of the modeling presented below relevant to our studies.

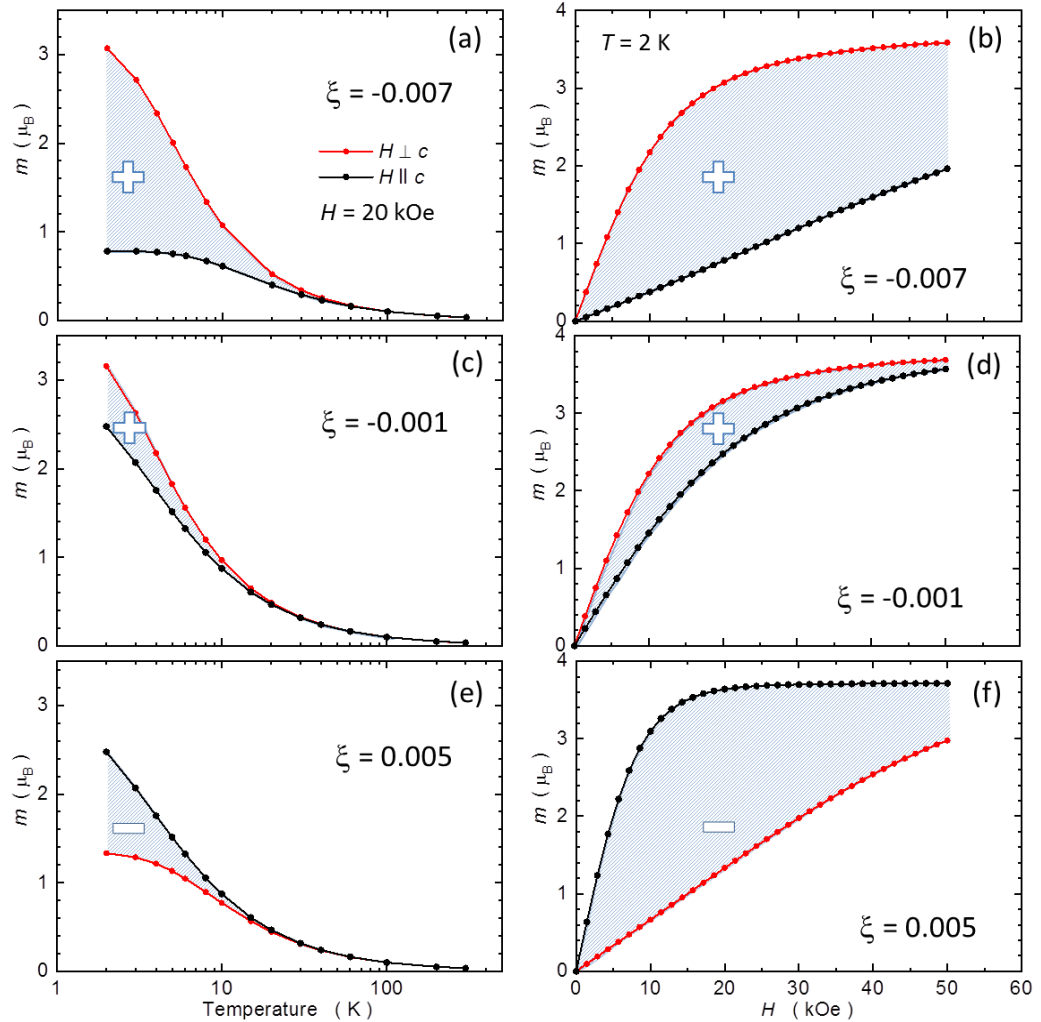

**Figure 3**. Single ion magnetic anisotropy for Mn$^{3+}$ ion in GaN calculated as a function of temperature $T$ at $H$ = 20 kOe [left panels (a),(c), and (e)], and as a function of magnetic field $H$ at $T$ = 2 K [right panels (b),(d), and (f)], for a model system of wurtzite GaN:Mn. The red symbols represent the $H \perp c$ orientation, the black ones the $H \parallel c$ one, where $c$ is the wurtzite hexagonal axis. The magnetic moment $m$ is expressed in Bohr magnetons. We define the positive magnetic anisotropy when $m(H \perp c) > m(H \parallel c)$. The corresponding signs are indicated in each panel. The key model parameter here is the magnitude of $\xi = c/a - (8/3)^{1/2}$, which quantify the relative magnitude of the trigonal deformation from the ideal wurtzite structure [39].

The results of the modeling of MA in GaN:Mn are collected in Figures 3 and summarized in Figure 4 (a). We have checked that qualitatively the results do not depend on the choice of the strength of $H$ from the range of fields available in commercial magnetometers. Therefore, for brevity we present only the results obtained for $H$ = 20 kOe, the same field as used to established experimental $\Delta m(T)$ in sapphire. The strength of the trigonal distortion can be quantified by a parameter $\xi = c/a - (8/3)^{1/2}$, i.e. by the relative magnitude of the trigonal deformation from the ideal wurtzite structure **[39]**. The pairs of panels (a)-(b), (c)-(d), and (e)-(f) in Figure 3 exemplify cases of $\xi$ = -0.007 (corresponding to a free standing GaN:Mn, i.e. a slightly compressed wurtzite crystal along its $c$-axis), $\xi$ = -0.001 (a nearly ideal wurtzite), and $\xi$ = 0.005, a uniaxially elongated GaN:Mn, respectively. The reversed strengths of the two branches of $m(H)$ in panel (f) with respect to the branches in panels (b) and (d) indicates the change of sign of the magnetic anisotropy for $\xi > 0$. Most importantly here, the same reversed strength is seen in panel (e). So, the change of the sign of MA is equally clearly reflected in the temperature domain. We now enumerate the differences between each pair of branches within the experimental ranges of $H$ and $T$ by calculating the corresponding areas, $A_H$ and $A_T$, respectively (indicated in Figure 3). The integration in $H$ domain is limited at $H_{max}$ = 50 kOe, the lower limits of $H$ available in MPMS SQUID magnetometers. We confirm that the results are qualitatively the same if $H_{max}$ = 70 kOe is adopted. The resulting dependency of $A_T$ on $A_H$ is plotted in Figure 4 (a). The points collected there show that indeed a quantification of the strength of MA on the account of $A_T$ yields qualitatively the same results as using measurements in the field domain. Let us point it out once again that the analysis of results in temperature domain does not provide with absolute values of MA, yet it correctly yields the sign of MA and is more experimentally efficient. In practice, the choice of $H$ should take into account both the expected magnitude of $\Delta m(T)$ and the noise level of the system. In the case reported here anything between 10 and 40 kOe proves satisfactory, since MA is sufficiently developed and the noise from the superconducting magnet remains within acceptable limits. The quality of the experimental evaluation deteriorates on moving to stronger fields - the magnitude of MA drops and SQUID magnetometer yields more noisy readings.

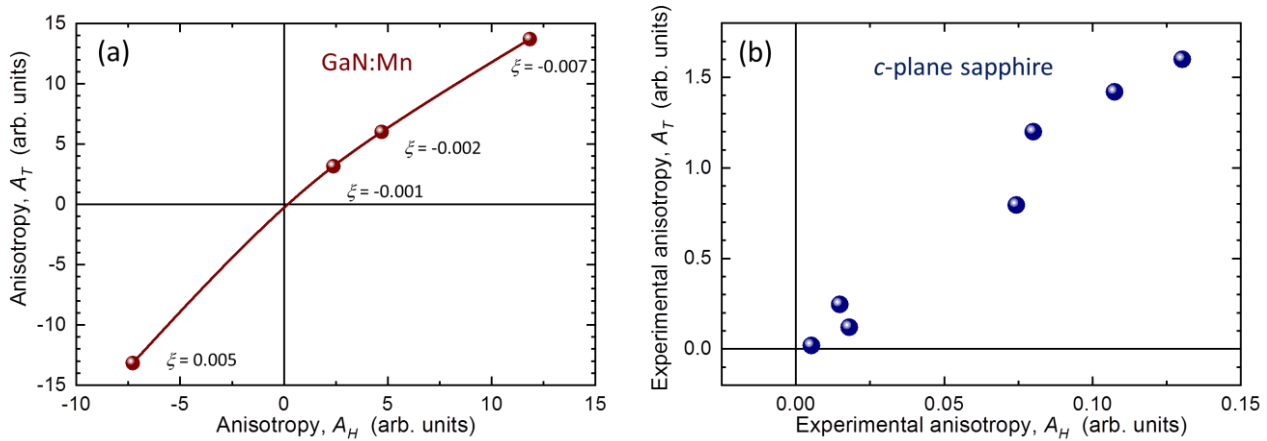

**Figure 4.** (**a**) Magnitudes $A_H$ and $A_T$ of the hatched areas between the branches of $m(T)$ and $m(H)$ presented in Figure 3. Their signs represents the orientation of the magnetic anisotropy, according to the labels in Figure 3 (a). The solid line is the guide for the eye. (**b**) The same as in (a) but calculated from experimental data for some of the sapphire samples tested in this study, including these presented in Figure 2. The sign conversion adopted here indicates that the positive magnitudes of $A_H$ and $A_T$ in panel (b) mean that the easy orientation of the experimental magnetic anisotropy in these $c$-plane sapphire samples is for the magnetic field applied perpendicularly to the sample plane ($H \parallel c$).

To conclude this part we plot in Figure 4 (b) the magnitudes of $A_T$ versus $A_H$ for some of the sapphire substrates investigated here. The points form a very similar pattern that is seen in the top-right quarter on panel (a). Firstly, it means that indeed the integration in $T$-domain can be used to quantification of MA in such materials like sapphire. Secondly, the points in panel (b) tend to bunch around (0, 0) and at higher values of $A_T$ and $A_H$. This indicates that in the whole range of the investigated $c$-plane sapphire substrates either there is practically no MA or a relatively strong MA develops at low temperatures. In this case the paramagnetic change of $m$ is greater for the perpendicular orientation of $H$ than when for the in plane configuration. It remains interesting to be seen whether other sapphire crystals can exert also a reversed MA, or the pattern observed here is a universal one for these commercial substrates.

## 5. Quantification of the strong paramagnetic component

Finally, we will attempt to shed some light on the origin of the strong Brillouin-like form of experimental $\Delta m(H)$, as shown in panels (b), (d), and (f) of Figure 2. The character of $\Delta m(H)$ reported there resembles greatly that of a single ion magnetic anisotropy, as presented in our numerical modeling, panels (b), (d) and (f) of Figure 3. However, nothing is known about the oxidation state and the exact coordination of these ions, so we cannot rely on the precision of the crystal filed model. Therefore, we resort to the classical Brillouin function $B_S(H, T)$:

$$B_S(x) = \frac{2S+1}{2S} coth\left(\frac{2S+1}{2S}x\right) - \frac{1}{2S} coth\left(\frac{1}{2S}x\right),$$

with $H$ and $T$ tied in $x$ by:

$$x = S\frac{g\mu_B H}{k_B T}.$$

to model the magnetic response of these ions. Here, $g = 2$ is the Land'e factor, $\mu_B$ is the Bohr magneton, and $k_B$ is the Boltzmann constant. However, we know that there exist another admixture to experimental $\Delta m(H)$, that one which dominates in Figure 3 (h). Since we cannot account numerically for this contribution we employ a technical trick outlined below. It allows a separation of the PM contribution from the other magnetic sources, usually exhibiting a FM-like form of their $m(H)$, like the response of FM nanocrystals embedded in the paramagnetic matrix **[13,49,61,62]**, or a form of the Langevin function exerted by superparamagnetic–like phase-separated mesoscopic volumes **[31]**.

In the first step of our approach we enumerate the experimental difference between the magnetic isotherms measured at 2 and 5 K, $\Delta m(H, \Delta T) = \Delta m(H, 2\,K) - \Delta m(H, 5\,K)$. The results are marked in the corresponding panels by magenta bullets. The main point here is that $\Delta m(H, \Delta T)$ is largely devoid of the non-PM components since the $T$-induced changes of $m(H)$ at low temperatures are the strongest for paramagnets. The other components largely cancel out. In the last step we fit to these $\Delta m(H, \Delta T)$ the difference of two Brillouin functions taken at the same temperatures as above. Namely, we fit $N_S g\mu_B S[B_S(H, 2\,K) - B_S(H, 5\,K)]$, where $N_S$ is the number of ions with spin $S$. Both $N_S$ and $S$ are the only adjustable parameters. The results of the fits are indicated by dashed lines, indicating that a very reasonable fit can be obtained in all three cases, that is when $\Delta m(H)$ is dominated by the PM contribution. The magnitudes of $N_S$ and $S$ are given in the corresponding panels. As said, these values cannot be treated as the exact numbers but we can safely conclude that the spin state of these ions is close to $S = 2$ (for $g = 2$) and that the number of these ions increases from $1 \times 10^{14}$ to $4 \times 10^{14}$ (their concentration changes from 1.3 to $6 \times 10^{16}$ cm$^{-3}$, respectively) on going from sample D to A1, as indicated by the increasing magnitudes of $\Delta m(H)$ along these three samples. We did not apply this procedure to data in Figure 2 (h), since these $\Delta m(H)$ cannot be approximated by classical Brillouin function. It is beyond the scope of this report to pursue the origin of this rather peculiar $\Delta m(H)$.

## 6. Conclusions

The simplified method of the assessment of the magnetic anisotropy in solid state substrates has been presented. Its applicability has been validated by magnetic investigation of common sapphire epi-ready wafers available on the market. It has been experimentally evidenced that in order to acquire a qualitatively correct information on the magnetic anisotropy of the material, instead of technically more cumbersome measurements in the magnetic field domain, one can resort only to temperature dependent studies. The presented experimental data and the theoretical modelling substantiate the fact that qualitatively correct information about of the strength and the sign of the magnetic anisotropy can be obtained by integration of the difference of the results of two temperature dependent measurements performed for two relevant orientations of the magnetic field. This allows for a more reliable, faster and economical efficient classification or selection of the material in possession, or to select the vendor with the most suitable substrates for the planned study. Very importantly, the method detailed here is practically immune to all the already recognized factors which mar the precision volume magnetometry performed in commercial SQUID magnetometers. This fact additionally saves experimental time and confusion and allows to obtain reliable outcome even by a novice in magnetometry.

Although the method has been tested and validated for c-plane sapphire wafers, the authors are confident that it can be applied to other types of sapphire substrates and crystalline solids like Si, GaAs, etc. for which the body of experimental evidences allows to define a convenient temperature range in which the normalization of the results obtained in different orientations of the magnetic field can be performed.

Concerning sapphire, our detailed study indicates that most likely all the *c*-plane sapphire substrates that have been available on the market do not exhibit purely diamagnetic response. They exhibit a net paramagnetic contribution of various strengths. We find that this contribution can be as strong as about 2% of the pure diamagnetic response of sapphire, rigorously evaluated here to be $\chi_{Sapphire}(300\ K) = -3.7(1) \times 10^{-7}$ emu/gOe. In absolute numbers this paramagnetism can bring as much as $2 \times 10^{-5}$ emu at 20 kOe, a number which easily exceeds the magnitudes of the magnetic responses seen in nanoscale structures, including 2D materials. This is the range of numbers that one should really reckon with and it indicates that the adequate magnetic testing of the substrate material should be mandatory in the magnetometric studies of miniscule nanostructures deposited on sapphire substrates. It has been evaluated that this paramagnetic component may be brought about by some $10^{16}$ cm$^{-3}$ transition metal ions. More importantly our study indicates that sapphire crystals exhibiting net paramagnetic responses exceeding about 2% of their diamagnetism do exert a magnetic anisotropy with the easy axis oriented along the *c* wurtzite axis of sapphire, that is being perpendicular to the plane of the *c*-plane substrates. Additionally, yet another paramagnetic-like contribution in sapphire has been identified, which is seen only when the concentration of the transition metal ions is very small. It exhibits a much more ferromagnetic like shape of its magnetization curves, but it is devoid of magnetic anisotropy. The origin of these spurious contributions to the diamagnetism of sapphire remains to be unraveled.

Our findings form an important message that this non-standard magnetism of commonly available substrates can significantly influence the results of precise magnetometry of nanoscale materials and that its existence of must be taken for granted by both industry and academia.

**Funding:** This study has been supported by the National Science Centre (Poland) through project OPUS 2018/31/B/ST3/03438.

**Appendix A**

In order to largely reduce the detrimental influence of the sample geometry on the results of the measurements of the magnetic anisotropy it is advisable to prepare the sample in a shape that has got the same $\gamma$ factor in both relevant orientation in the magnetometer. Again, the problem is in the large magnitude of the signal of the substrate. It can be as large as about -10$^{-3}$ emu in 70 kOe, so a change of the orientation of $5 \times 5$ mm$^2$ specimen from the in plane into the perpendicular orientation causes changes of the coupling to the SQUID pick-up coils from 0.983 to 1.034 **[26]**. Accordingly, even for an isotropic material the signal reported by MPMS SQUID magnetometer will change by 5%, which is about $5 \times 10^{-5}$ emu in this example.

To minimize this massively detrimental effect we are preparing our samples by catting the material into narrow strips of approximately $1.3 \times 5$ mm$^2$. Stacking four of them on top of another forms a cuboid of approximately squared cross section $1.3 \times 1.3$ mm$^2$ **[39]**. Now, any rotation by 90° along the long side of the cuboid facilitates the change of the orientation between the in plane and the perpendicular one with marginally weak changes of the $\gamma$ coupling factor, thus largely preserving the correspondence of the data collected in both orientations. The experimental realization is exemplified in Figure A1.

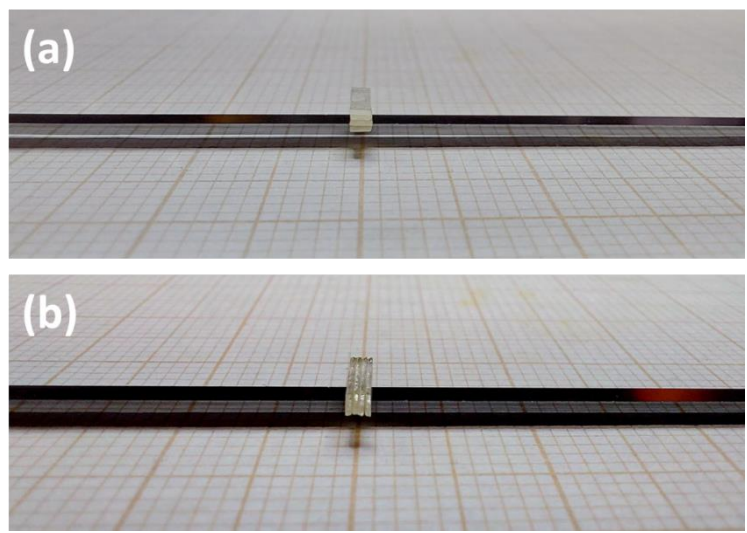

**Figure A1.** An example of the modification of a 5 × 5 mm² piece of 0.3 mm thick sapphire sample into a symmetrical assembly of approx. 1.3 × 1.3 × 5 mm³. The sample is mounted onto a long Si stick serving as the final part of the QD MPMS typical sample holder for RSO measurements. (**a**) The sample is mounted for the in plane measurements ($H$ is parallel to the surface of the original piece), and (**b**) the same sample is mounted for the perpendicular orientation.